\documentclass{article}
\usepackage{arxiv}

\usepackage[utf8x]{inputenc}
\usepackage[T1]{fontenc}
\usepackage{cite}
\usepackage[cmex10]{amsmath}
\usepackage{amssymb} 
\usepackage{mathrsfs, color} 

\usepackage{graphicx,psfrag}
\usepackage{booktabs} 

\DeclareMathOperator*{\argmin}{arg\;min}

\DeclareMathOperator*{\dive}{div}

\begin{document}
	
\title{Noise-specific denoising method with applications to high-frequency ultrasonic images}
	
\author{
  Gonzalo~D.~Maso~Talou \\
  Auckland Bioengineering Institute\\
  The University of Auckland\\
  Auckland, New Zealand\\
  \texttt{g.masotalou@auckland.ac.nz} \\
   \And
 Pablo~J.~Blanco \\
  National Laboratory for Scientific Computing (LNCC)\\
  National Institute of Science and Technology \\
  in Medicine Assisted by Scientific Computing (INCT/MACC)\\
  Petr\'opolis, Brazil \\
  \texttt{pjblanco@lncc.br} \\
}
	
		

	
\maketitle
	
\begin{abstract}
        Denoising is of utmost importance for the visualization and processing of images featuring low signal-to-noise ratio. Total variation methods are among the most popular techniques to perform this task improving the signal-to-noise ratio while preserving coherent intensity discontinuities. 
        In this work, a novel method, termed maximum likelihood data, is proposed, endowing the total variation formulation with the capability to deal with noise-specific models and pre-processing stages for a certain image of interest. To do this, the data fidelity term is modified by means of a maximum likelihood estimator between the original and the denoised image.
		To assess the improvements of the proposed method with respect to the total variation formulation, we study the denoising of high-frequency ultrasonic images on in-silico and in-vivo setups. The proposed method delivered a better contrast, preservation and localization of the structures while diminishing the intensity bias of the total variation formulation for the multiplicative noise.
		The enhancement of medical images through denoising helps to improve the outcome of subsequently applied image processing such as registration and segmentation procedures. 
\end{abstract}
	
	
\section{Introduction}

Image denoising is an essential task for visualization, specifically in images featuring low signal-to-noise ratio (SNR). The image analysis and processing, such as registration, segmentation, motion tracking or pattern recognition, pose serious challenges in the context of these poor quality images. Usually, different approaches are applied according to the characteristics of the acquisition sensor or its acquisition setup. 

As proposed in \cite{jain2016survey}, most of the state-of-the-art techniques for edge-preserving denoising can be categorized as: i) Median filtering \cite{jain1989fundamentals,gonzalez2012digital}; ii) Total variation based filtering \cite{Rudin1992,Chambolle2004,gilboa2008nonlocal,bredies2010total,tang2016edge}; iii) Anisotropic diffusion \cite{krissian2007oriented, perona1990scale, black1998robust,weickert1998efficient}; iv) Bilateral filtering \cite{tomasi1998bilateral,durand2002fast,gastal2012adaptive,chaudhury2013acceleration,mozerov2015global}; v) Non-local means (NLM) based filtering \cite{buades2005non,coupe2008optimized,tasdizen2008principal,maggioni2013nonlocal}; and vi) Wavelet transform based filtering \cite{donoho1994ideal,donoho1995adapting,chang2000adaptive,qiu2013llsure}. In this work, we are focused in total variation filters presenting how to extend its formulation to consider specific noise models.
	
Particularly, the ROF method \cite{Rudin1992} offers a general purpose solution for different kinds of images, based on the assumption that a high absolute gradient over the image (i.e. total variation) represents spurious variations caused by noise. In such method, the minimization of the total variation is complemented with a data fidelity term that penalizes the differences between the original image and the denoised one, allowing the conservation of relevant structures in the image. Thus, a linear combination of these two terms leads to the target denoised image.

Posterior developments in ROF-like formulations have been focused in the enhancement of total variation term with higher order terms. In \cite{chambolle1997image}, a inf-convolution of two or more convex variation terms is proposed. In \cite{chan2000high}, an additional second order term is added to the formulation, such that the total variations contribution are dumped as the first order term identifies image edges. These approaches diminish the appearance of stair-case artifacts, while preserving the sharpness of the structures in the images. Furthermore, \cite{bredies2010total} proposes the total
generalized bounded variation (or simply TGV), presenting a polynomial regularizer that balances the first $k$-th order variations. It is shown how the addition of higher order variations allows the data to fit higher order patterns to avoid undesirable piece-wise constant (in first order) or linear (second order) interpolations across restored regions. Even an extension of this approach was recently proposed in \cite{kongskov2017directional} to weigh the contribution of TGV among different directions to obtain anisotropic restoration patterns.

Conversely, the data fidelity term is frequently chosen to be the absolute difference between the original image and the denoised one, and variants of the data fidelity term in the ROF formulation have been seldom explored \cite{krissian2005speckle, le2007variational, rodriguez2008efficient}.

In this work, we propose a generalization of such data fidelity term using an inverse maximum likelihood estimator that captures the nuances of the image noise, according to a specific probabilistic model, enhancing the data preservation after denoising. Particularly, we present the classic ROF data fidelity term as a particular case of our approach when a zero mean Gaussian noise is considered. Moreover, we derive an appropriate formulation for high-frequency ultrasound images to exemplify the use of the methodology to obtain a noise specific data fidelity term. Here, we only make use of the $L_1$ total variation term in the formulations to perform straightforward comparisons and assess the advantages of a tailored data fidelity term. Evidently, the use of more sophisticated total variation terms (such as those proposed in \cite{bredies2010total,tang2016edge}) could lead to further improvements in the outcome of the denoising process.

The high-frequency ultrasonic case is appealing to study the advantages of a more detailed data fidelity term in the ROF formulation because the noise can be modeled using a non-conventional representation (log-compressed and multiplicative noise \cite{Michailovich2006}) and a non-symmetric, neither Gaussian distribution (generalized gamma distribution \cite{raju2002statistics}). Several approaches were proposed in the literature for denoising of these images \cite{Michailovich2006,rizi2012noise,lazrag2012despeckling,mitra2015wavelet,soorajkumar2016fourth,wen2016nonlocal} because of its relevance in the medical field. As a low SNR image, the reduction of the speckle noise is essential to enhance its post-processing and expert examination.

Next, we introduce the proposed method by deriving it from a TV-L1 formulation (Section~\ref{sec:Despeckle}). Particularly, the formulations for generalized gamma and normal noise distributions are described in Section~\ref{sec:MLDGammaNoise} and Appendix~\ref{sec:MLDGaussian}, respectively. Performance and illustrative applications of the method are analyzed in Section~\ref{sec:results} for in-silico and in-vivo ultrasonic medical images.
	
\section{Denoising methods} \label{sec:Despeckle}

The proposed denoising method is developed targeting the specific noise distributions that affects the image of interest. For the formulation of the data fidelity term, we propose a dissimilarity function based on the maximum likelihood estimator (MLE) on top of the probability model postulated for the noise. To describe this process, we instantiate our method for log-compressed images with gamma generalized noise that corresponds to the expected noise model for high-frequency ultrasonic images. In what follows, we described the formulations of the TV-L1 and the proposed maximum likelihood data (MLD) for image denoising.
	
\subsection{Total variation method} \label{sec:TVL1}
	
As proposed in \cite{Rudin1992}, the total variation method allows the image denoising for a wide spectrum of noise distributions. The method yields a denoised image by minimizing the total variation with respect to the original image. Particularly, the use of the $L_1$ instead of $L_2$ norm over the total variation term (denoted as TV-L1) preserves better the coherent discontinuities (e.g. edges of a structure) while maintaining the denoising level. To formally define the TV-L1 method, let us call $I(x,y)$ to the original image and $J(x,y)$ to the denoised image, then $J(x,y)$ is estimated as follows
\begin{equation}
	J = \argmin_{ \tilde{J} } \int_{\Omega} \Big( \lvert I(x,y) - \tilde{J}(x,y) \rvert + \alpha \lvert \nabla \tilde{J}(x,y) \rvert \Big) \, d\Omega,
	\label{eq:TVL1}
\end{equation}    
where $\Omega$ is the image domain. The functional defined in \eqref{eq:TVL1} is composed by a data term that measures the absolute difference between the images and by a regularization term that imposes the continuity of the intensity in the denoised image. The $\alpha$ parameter is key for the expected outcome of the method, small values of $\alpha$ are unable to provide the adequate denoising effect, while large values do not preserve image structures. For this reason there is a compromise in the choice of $\alpha$, having to be small enough to preserve the structures of interest in the image and large enough to properly deal with the noise. Noteworthy, in this particular method, parameter $\alpha$ has no physical nor experimental association.

To efficiently minimize \eqref{eq:TVL1}, it is used the proposed Primal-Dual method by \cite{Chan1999} in which a saddle point formulation of the functional is adopted. This method improves global convergence behavior in comparison to the primal
Newton's method. Convergence and numerical analysis for this approach are presented in \cite{Chambolle2004}. 

As presented in \eqref{eq:TVL1}, the data fidelity term searches for the closest $J$ to $I$ in the sense of the $L_1$-norm such that satisfies the spatial smoothness imposed by the regularization term. Next, we propose an alternative to the $L_1$-norm difference to establish a correspondence between $I$ and $J$, exploiting the knowledge of the noise probability distribution.
	
\subsection{Maximum likelihood datum method} \label{sec:MLD}

The data fidelity term in the TV-L1 formulation acts as a spring with a force magnitude proportional to $\lvert I-J \rvert$ and stiffness parameter equals to $\alpha^{-1}$. If the noise distribution is Gaussian and zero-mean, the data term provides an adequate performance since it penalizes equally positive and negative differences (symmetry of the normal distribution) and penalizes less the most probable noise (close to zeros, since it is zero-mean) allowing the regularization term to smooth such regions. Conversely, noise with non-zero mean distributions will yield an intensity bias and non-symmetrical distributions will not be treated properly.

Thus, we propose a new formulation that penalizes the discrepancy between $J$ and $I$ by the noise probability instead of the noise modulus. The so-called maximum likelihood data (MLD) method estimates the denoised image $J(x,y)$ as follows
\begin{equation}
	J = \argmin_{ \tilde{J} } \int_{\Omega} \Big( -c(I(x,y), \tilde{J}(x,y)) + \alpha \lvert \nabla \tilde{J}(x,y) \rvert \Big) \, d\Omega,
	\label{eq:MVL}
\end{equation}
where $c$ is the maximum likelihood estimator (MLE) for images $\tilde{J}$ and $I$. Note that the data term uses the negative MLE to measure the probabilistic discrepancies between images in a proper manner. For the derivation of the $c$ function for a particular type of images, an appropriate noise model --noise probability distribution and pre-processing model-- is required. The data term with $L_2$-norm from ROF-like methods corresponds to the MLD method when normally distributed and zero-mean noise without pre-processing is assumed (see Appendix~\ref{sec:MLDGaussian}). In the following section, we describe the construction of the MLD functional considering the pre-processing model (log-compression of the images) and the noise distribution (generalized gamma noise).

\subsection{\textnormal{MLD} for high-frequency ultrasonic images}\label{sec:MLDGammaNoise}

According to \cite{raju2002statistics}, the generalized gamma (GG) distribution appears to be a reasonable probabilistic model for the noise present in high-frequency ultrasonic images. To derive an appropriate maximum likelihood estimator, let us define $u$, an image with GG distributed noise, and $v$, the noiseless version of $u$. The noise model for an ultrasonic image is given by
\begin{equation}
	u(x,y) = v(x,y) \; \varepsilon(x,y)
\end{equation}
where $\varepsilon(x,y)$ is a random variable with generalized gamma distribution that models the speckle noise. To ease the notation, we drop the indexes from the images, and their operations must be interpreted as pixel-wise. For visualization purpose, the ultrasonic images undergo log-compression of its dynamic range, then
\begin{eqnarray}
	\ln \big( u \big) &=& \ln \big( v \, \varepsilon \big) \nonumber\\
	\ln \big( u \big) &=& \ln \big( v \big) + \ln \big(\varepsilon \big) \nonumber\\
	I &=& J + \tilde \varepsilon
	\label{eq:noiseModelTwoImages}
\end{eqnarray}
where $I$ and $J$ are the log-compressed versions of $u$ and $v$, respectively, and the random variable $\tilde \varepsilon$ is distributed with probability density function (PDF) $P_{\tilde \varepsilon}(\tilde \varepsilon)$ derived from the GG distribution $P_{\varepsilon}(\varepsilon)$. Random variables $\varepsilon$ and $\tilde \varepsilon$ are related as $\tilde \varepsilon=\ln \varepsilon$, and the PDFs satisfy
\begin{eqnarray*}
	P_{\tilde \varepsilon}(\tilde \varepsilon) d\tilde \varepsilon &=& P_{\varepsilon}(\varepsilon) d\varepsilon \\
	P_{\tilde \varepsilon}(\tilde \varepsilon) &=& P_{\varepsilon}(\varepsilon) \frac{d\varepsilon}{d\tilde \varepsilon}
\end{eqnarray*}
By changing variables, $\varepsilon = e^{\tilde \varepsilon}$ it follows
\begin{eqnarray*}
	P_{\tilde \varepsilon}(\tilde \varepsilon) &=& P_{\varepsilon}(e^{\tilde \varepsilon}) e^{\tilde \varepsilon}\\
	&=& \frac{\gamma}{\delta^{\gamma \, \nu}\, \Gamma(\nu)} \, (e^{\tilde \varepsilon})^{\gamma \, \nu - 1} \, e^{-\big(\frac{e^{\tilde \varepsilon}}{\delta}\big)^\gamma} e^{\tilde \varepsilon}.
\end{eqnarray*}
After some algebra, the log-compressed generalized gamma PDF results
\begin{equation}
	P_{\tilde \varepsilon}(\tilde \varepsilon) = \frac{\gamma}{\Gamma(\nu)} e^{\gamma \nu (\tilde \varepsilon - \ln \delta) - e^{\gamma(\tilde \varepsilon - \ln \delta)}}, \quad \gamma,\delta,\nu > 0.
	\label{eq:PDFGGlogcomp}
\end{equation}
	
Using this speckle noise model, we construct the associated MLE as
\begin{eqnarray}
	e(I,J) &=& \int_{\Omega} \ln \bigg[ P_{\tilde \varepsilon}(I-J) \bigg] \nonumber\\
	&=& \int_{\Omega} \ln \frac{\gamma}{\Gamma(\nu)} e^{\gamma \nu (I - J - \ln \delta) - e^{\gamma(I - J - \ln \delta)}}\nonumber\\
	&=& C_1 \int_{\Omega} \bigg[ \gamma \nu (I - J) - \frac{1}{\delta^{\gamma}} e^{\gamma(I - J)} \bigg] \nonumber\\
	&&\mbox{with } C_1 = \lvert \Omega \rvert \bigg(\ln \frac{\gamma}{\Gamma(\nu)} - \gamma \nu \ln \delta \bigg) 
	\label{eq:MLEforLogGG}
\end{eqnarray}
where $\Omega$ is the image domain. The constant contribution $C_1$ is discarded because it does not affect the denoising process. The MLE for high-frequency ultrasound is given by
\begin{equation}
	c(I, J) = \gamma \nu (I - J) - \frac{1}{\delta^{\gamma}} e^{\gamma(I - J)}.
	\label{eq:DenoisingGGC}
\end{equation}
which is replaced in \eqref{eq:MVL} to obtain the MLD formulation for our specific noise distribution, i.e.,
\begin{equation}
	J = \argmin_{ \tilde{J} } \int_{\Omega} \Big( - \gamma \nu (I - \tilde{J}) + \frac{1}{\delta^{\gamma}} e^{\gamma(I - \tilde{J}}) + \alpha \lvert \nabla \tilde{J} \rvert \Big) \, d\Omega.
	\label{eq:MLD}
\end{equation} 
\begin{figure}
	\centering
	\includegraphics{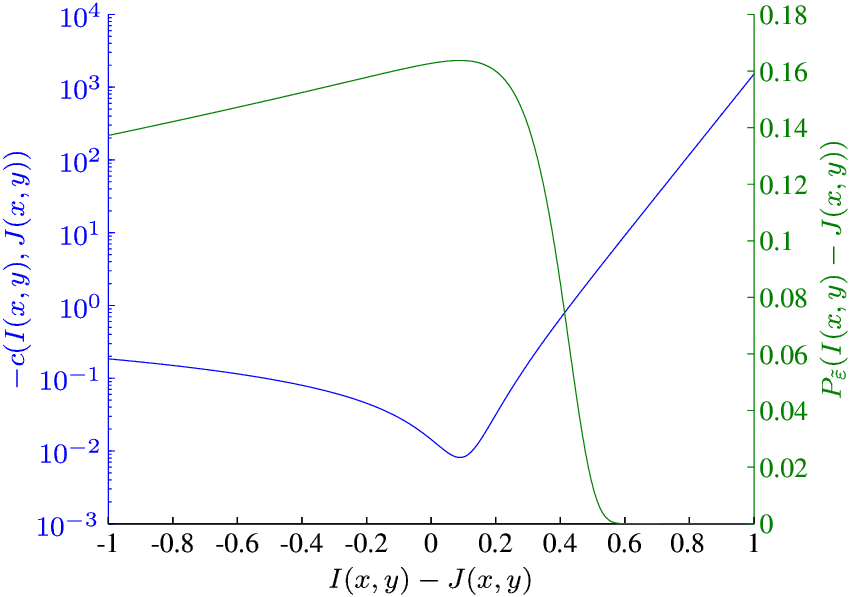}
	\caption{Probability density function for log-compressed noise in high-frequency ultrasonic images (green line) versus the data term value in \eqref{eq:MLD} (blue line). Parameters used both functions are $\gamma=12.75$, $\nu=0.014$ and $\delta=1.53$.}
	\label{fig:ComparisonGGDistvsMLE}
\end{figure}
As seen in Figure~\ref{fig:ComparisonGGDistvsMLE}, the higher the probability of the difference between the original and the denoised image ($P_{\tilde \varepsilon}$), the smaller the penalization value in the data term, enabling its correction by the regularization term. That is the same desirable behavior featured by the TV-L1 method for Gaussian noise distributions. Here, this property in MLD is granted by construction, for any noise distribution. 

The Euler-Lagrange equations associated to the minimization problem of cost functional \eqref{eq:MLD} are derived in detail in Appendix~\eqref{sec:NecSufConditionsAndELEq} (demonstrating uniqueness and existence of the solution) and are the following
\begin{eqnarray}%
	\gamma \nu - \frac{\gamma}{\delta^\gamma} e^{\gamma \, (I - J)}	- \alpha \dive \bigg( \frac{\nabla J}{\lvert \nabla J \rvert} \bigg) &= 0,& \mbox{ in } \Omega \label{eq:MLVEulerLagrange}\\ 
	\nabla J \cdot \mathbf n &= 0,&\mbox{ in } \partial \Omega
\end{eqnarray}
where $\partial \Omega$ is the boundary of the image domain, whose unit outward normal vector is $\mathbf n$. A finite difference approach is employed to discretize the differential operators and a fixed-point iteration scheme with sub-relaxation is used to deal with non-linearities. Thus, the linearized problem in discrete form reads
\begin{equation}
	\begin{split}
	J^{n+1}_{x,y} = \beta \, J^{n}_{x,y} + (1 - \beta) \dfrac{ \gamma \nu - \frac{\gamma}{\delta^\gamma} e^{\gamma \, (I_{x,y} - J^n_{x,y})} + \alpha f^n_{x,y}}{\alpha s^n_{x,y}}
	\end{split}
\end{equation}
where $f^n_{x,y},s^n_{x,y}$ are defined as
\begin{equation}
	\begin{split}
	f^{n}_{x,y} =& - \frac{1}{\Delta x} \bigg( \frac{J^n_{x+1,y}}{\lvert G^n_{x+\frac{1}{2},y}\rvert} + \frac{J^n_{x-1,y}}{\lvert G^n_{x-\frac{1}{2},y}\rvert}\bigg)\\
				&- \frac{1}{\Delta y} \bigg( \frac{J^n_{x,y+1}}{\lvert G^n_{x,y+\frac{1}{2}}\rvert} + \frac{J^n_{x,y-1}}{\lvert G^n_{x,y-\frac{1}{2}}\rvert}\bigg)
	\end{split}
\end{equation}
\begin{equation}
	\begin{split}
	s^{n}_{x,y} =& - \frac{1}{\Delta x} \bigg( \frac{1}{\lvert G^n_{x+\frac{1}{2},y}\rvert} + \frac{1}{\lvert G^n_{x-\frac{1}{2},y}\rvert}\bigg)\\
				&- \frac{1}{\Delta y} \bigg( \frac{1}{\lvert G^n_{x,y+\frac{1}{2}}\rvert} + \frac{1}{\lvert G^n_{x,y-\frac{1}{2}}\rvert}\bigg).
	\end{split}
\end{equation}
The discrete gradient operator at the pixels edges is approximated by the following finite difference estimates
\begin{multline}
	G^n_{x+\frac{1}{2},y} = \\
	\bigg(\frac{J^n_{x+1,y}-J^n_{x,y}}{\Delta x},	\frac{J^n_{x+1,y+1} + J^n_{x,y+1} - J^n_{x+1,y-1} - J^n_{x,y-1}}{4 \Delta y} \bigg)\\
\end{multline}
\begin{multline}
	G^n_{x,y+\frac{1}{2}} = \\
	\bigg(\frac{J^n_{x+1,y+1} + J^n_{x+1,y} - J^n_{x-1,y+1} - J^n_{x-1,y}}{4 \Delta x} ,	\frac{J^n_{x,y+1} - J^n_{x,y}}{\Delta y}\bigg).
\end{multline}

\section{Results} \label{sec:results}

For both in-silico and in-vivo cases, the dynamic range of the images was normalized to $[0,1]$ interval using double precision for their representation and processing.

\subsection{Error quantification}

Three error functions were employed to quantify the different features of the denoising process: (i) mean intensity bias; (ii) a posteriori noise dispersion; and (iii) edge pattern recovery. 

The mean intensity bias due to denoising process is measured by a normalized $L_2$-norm of the differences between $J$ and a reference noiseless image $I^{\mathrm{ref}}$, i.e.,
\begin{equation}
\varepsilon_B = \sqrt{ \dfrac{\sum_{(x,y)\in \Omega}(I^{\mathrm{ref}}(x,y)-J(x,y))^2}{\sum_{(x,y)\in \Omega} (I^{\mathrm{ref}}(x,y))^2} }.\label{eq:eps_B}
\end{equation}

The a posteriori noise dispersion is estimated as the standard deviation of the pixel-wise differences between $J$ and $I^{\mathrm{ref}}$ in the following manner
\begin{equation}
\varepsilon_D = \sqrt{\dfrac{\sum_{(x,y)\in \Omega}(I^{\mathrm{ref}}(x,y)-J(x,y) - \mu_D)^2}{\lvert\Omega\rvert}}\label{eq:eps_D}
\end{equation}
where $\mu_D$ is mean value of $I^{\mathrm{ref}}-J$ and $\lvert \Omega \rvert$ is number of pixels in $\Omega$. Note that this measure quantifies the noise after denoising (i.e., $I^{\mathrm{ref}} - J$) correcting the intensity bias by subtracting $\mu_D$. 

Lastly, the success for the edge pattern recovery is measured by the function proposed in \cite{Michailovich2006}, given by
\begin{equation}
\varepsilon_E = \dfrac{\sum_{(x,y)\in \Omega} (\Delta I^{\mathrm{ref}}(x,y) \Delta J(x,y))^2}{ \sqrt{ \sum_{(x,y)\in \Omega} (\Delta I^{\mathrm{ref}}(x,y))^2} \, \sqrt{\sum_{(x,y)\in \Omega} (\Delta J(x,y))^2}}\label{eq:eps_E}
\end{equation}
where $\Delta$ is the discrete Laplace operator. As the Laplace operator is an edge detector for images, this function quantifies the correlation between the edges of $I^{\mathrm{ref}}$ and $J$.

\subsection{In-silico experiment}

\begin{figure*}
	\centering
	\includegraphics[width=5cm]{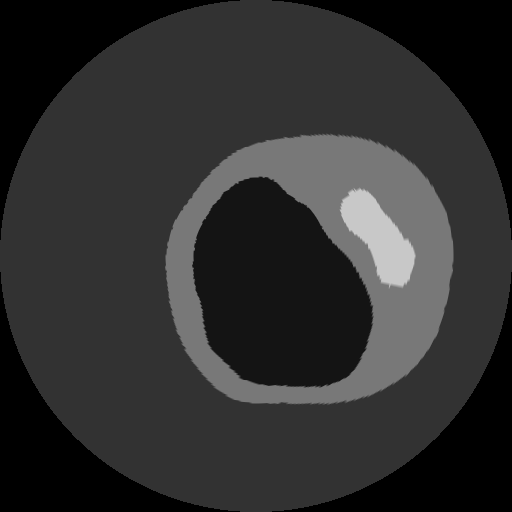}
	\includegraphics[width=5cm]{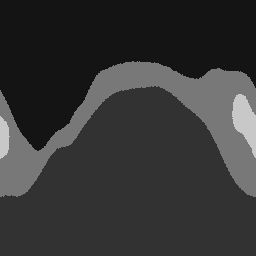}
	\includegraphics[width=5cm]{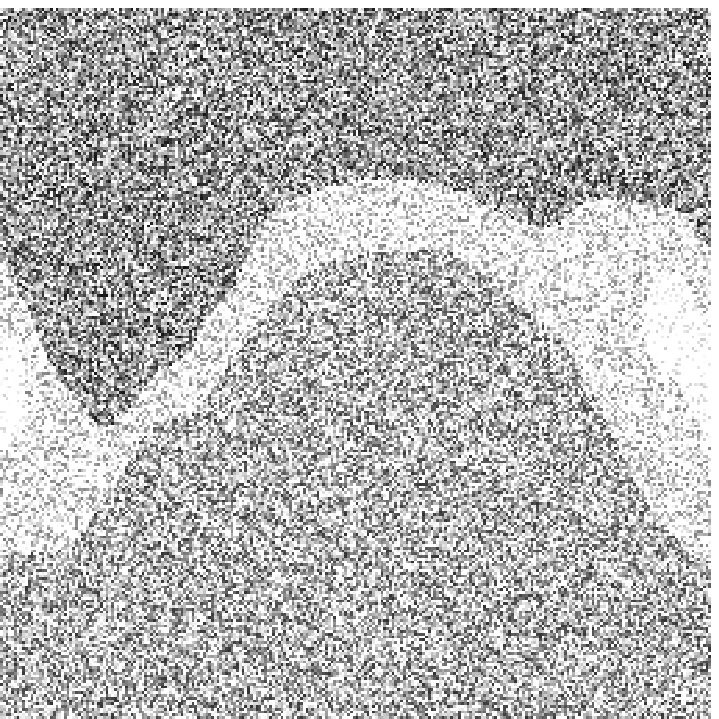}
	\caption{In-silico experiment modeling a cross-section arterial vessel in polar coordinates mimicking the acquisition of an IVUS transducer: (left) In-silico image in Cartesian coordinates; (center) In-silico image in polar coordinates (transducer native coordinates); (right) In-silico image with multiplicative GG distributed noise generated with parameters $\gamma = \nu = \delta = 1.5$.}
	\label{fig:InsilicoExperiment}
\end{figure*}

To assess the enhancements of the MLD with respect to TV-L1, we created an in-silico image mimicking an IVUS acquisition of a vessel cross-section with an atherosclerotic lesion (see Figure~\ref{fig:InsilicoExperiment}). The geometry and intensities for each structure in the image (lumen, intima, calcium and adventitia) were chosen inspired in in-vivo images. The noise was generated using the log-compressed generalized gamma PDF (see \eqref{eq:PDFGGlogcomp}). Then, we searched for the parameters of each method (MLD and TV-L1) that optimized the denoising process in the sense of the error functions defined in the previous section. To evaluate such error functions, the ground truth image (i.e. the in-silico image without noise) is used as $I^{\mathrm{ref}}$.

For each error function, a brute force strategy was applied to find the optimal parameters that reduced the error, i.e., a discretization of the parameter space of the methods was performed and the optimum set of parameters was then found. The discretization for the MLD parameter space was 
\begin{eqnarray}
\mathcal P^{\mathrm{MLD}} &=& \{ (\alpha, \gamma, \nu, \delta); \; \alpha \in \mathcal A, \; \gamma,\nu,\delta \in \mathcal B \} \\
\mathcal A &=& \{ 0.5, 0.75, 1, 2\} \nonumber\\
\mathcal B &=& \{ 0.1 \cdot i, \;i=1,2,\ldots,20\}\nonumber
\end{eqnarray}
and the discretization for the TV-L1 parameter space was
\begin{equation}
\mathcal P^{\mathrm{TV-L1}} = \{ \alpha; \; \alpha \in \{ 0.01 \cdot i, \; i=1,2,\ldots 100 \} \}.
\end{equation}

\begin{figure*}
	\centering
	\includegraphics[width=4.5cm]{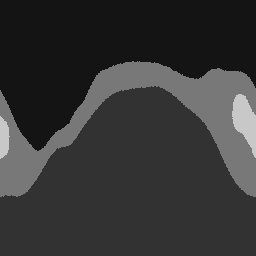}
	\includegraphics[width=4.5cm]{figures/fig2_InsilicoNoise.eps} \vspace*{0.1 cm}\\
	\includegraphics[width=4.5cm]{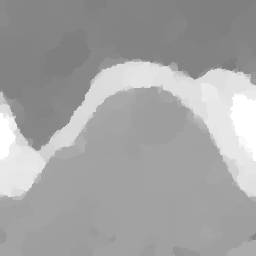}
	\includegraphics[width=4.5cm]{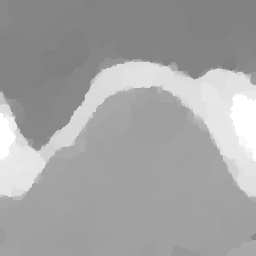}
	\includegraphics[width=4.5cm]{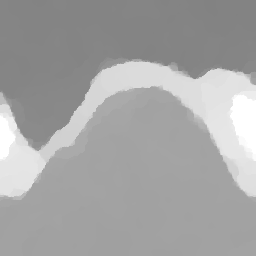} \vspace*{0.1 cm}\\
	\includegraphics[width=4.5cm]{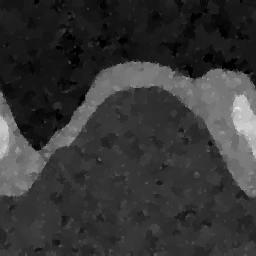}
	\includegraphics[width=4.5cm]{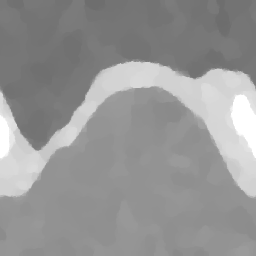}
	\includegraphics[width=4.5cm]{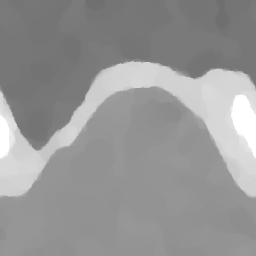} \vspace*{0.1 cm}
	\caption{Denoised in-silico experiment using the TV-L1 and MLD methods with optimal parameters (in the sense of error functions \eqref{eq:eps_B}, \eqref{eq:eps_D} and \eqref{eq:eps_E}). (Top) In-silico image without and with noise; (middle, from left to right) image denoised with TV-L1 using optimal parameters for $\varepsilon_B$, $\varepsilon_D$ and $\varepsilon_E$ error functions, respectively; (bottom, from left to right) image denoised with MLD using optimal parameters for $\varepsilon_B$, $\varepsilon_D$ and $\varepsilon_E$ error functions, respectively. Denoising parameters are reported in Table~\ref{tab:ErrorOpt}.} 
	\label{fig:ErrorOpt}
\end{figure*}

The results of such optimization processes are presented in Table \ref{tab:ErrorOpt} and Figure \ref{fig:ErrorOpt}. The MLD outperforms TV-L1 in terms of the three error functions when optimal parameters for both methods are employed. The TV-L1 presents a clear intensity bias as anticipated in Section~\ref{sec:MLD}, because its data term is minimized closer to the local-mean value in each region. In the case of the MLD, the data term is minimized to an intensity with low probabilities to feature the observed noise, which reduces the bias when appropriate parameters (noise distribution) are used. In qualitative terms, it is observed that the MLD with higher $\varepsilon_E$ renders the most similar image with respect to the ground truth in terms of intensity homogeneity, clear edges and less noise artifacts. In turn, the less intensity bias is obtained for the denoised image with the lowest $\varepsilon_B$ because it is closer to the true noise statistical parameters ($\gamma = \nu = \delta = 1.5$). 

Note that MLD delivered better results even using a less refined parameter space in comparison with TV-L1 (parameters in $\mathcal P^{\mathrm{MLD}}$ were less refined than the parameters in $\mathcal P^{\mathrm{TV-L1}}$), emphasizing that no elaborated parametrization setup is needed to obtain an improved outcome when the data fidelity term is appropriately modeled.

\begin{table*}
\centering
\begin{tabular}{l|cc|ccccc}
		\toprule
		Error				&	\multicolumn{2}{c|}{TV-L1}						&	\multicolumn{5}{c}{MLD} \\
		function			&	$\alpha$		&	Error value		&	$\alpha$	&	$\gamma$	&	$\nu$	&	$\delta$	&	Error value	\\
		\midrule
		$\varepsilon_B$		&	$0.80$			&	$1.83$						&	$0.5$		&	$1.4$		&	$1.4$	&	$1.3$		&	$0.04$						\\
		$\varepsilon_D$		&	$0.74$			&	$3.17 \cdot 10^{-2}$		&	$0.75$		&	$0.7$		&	$1.0$	&	$1.6$		&	$2.66 \cdot 10^{-2}$		\\
		$\varepsilon_E$		&	$0.58$			&	$0.36$						&	$0.5$		&	$0.8$		&	$0.8$	&	$1.3$		&	$0.51$						\\
		\bottomrule		
\end{tabular}
\vspace*{0.05cm}
\caption{Measures obtained with the error functions for the methods TV-L1 and MLD in the in-silico experiment. Also, optimized parameters in each case are reported. Error functions $\varepsilon_B$ and $\varepsilon_D$, should be close to zero, whereas error function $\varepsilon_E$ should be close to one.}
\label{tab:ErrorOpt}
\end{table*}
	
\subsection{In-vivo IVUS study}

\begin{figure*}
\centering
\includegraphics[height=3.cm]{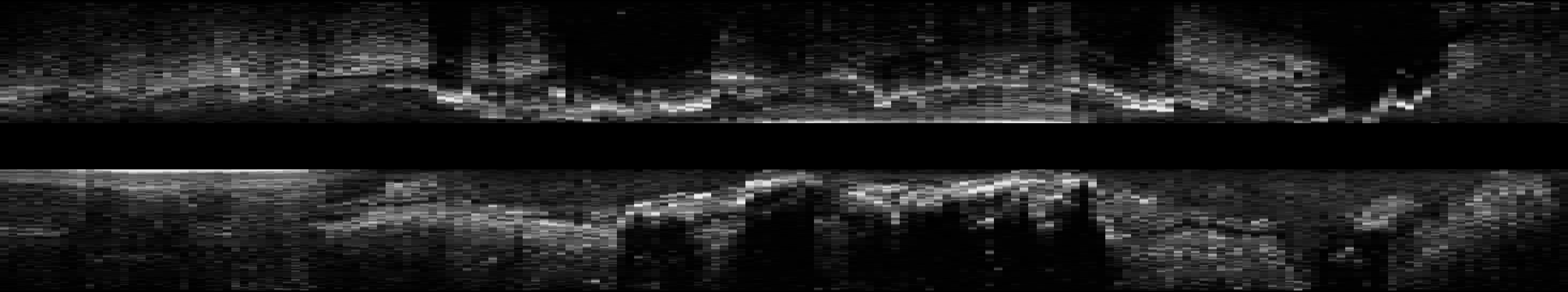}\\
\vspace*{0.05cm}
\includegraphics[height=3.cm]{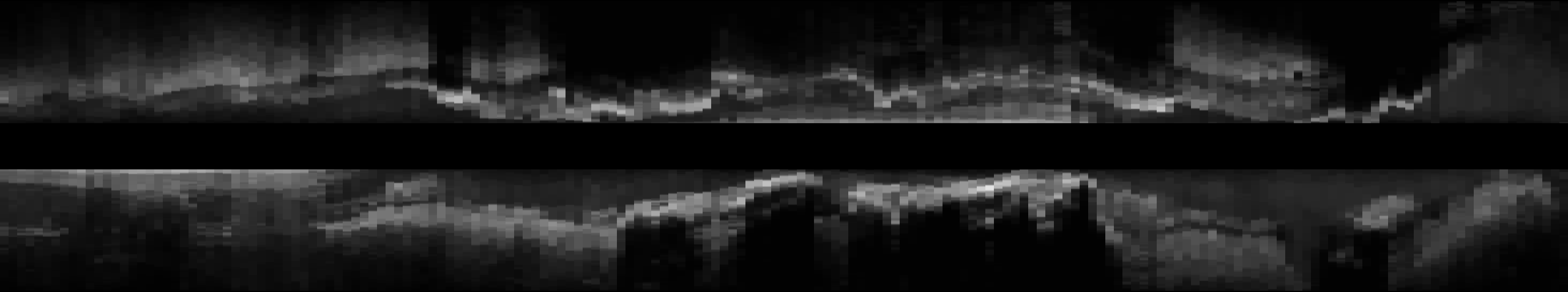}\\
\vspace*{0.05cm}
\includegraphics[height=3.cm]{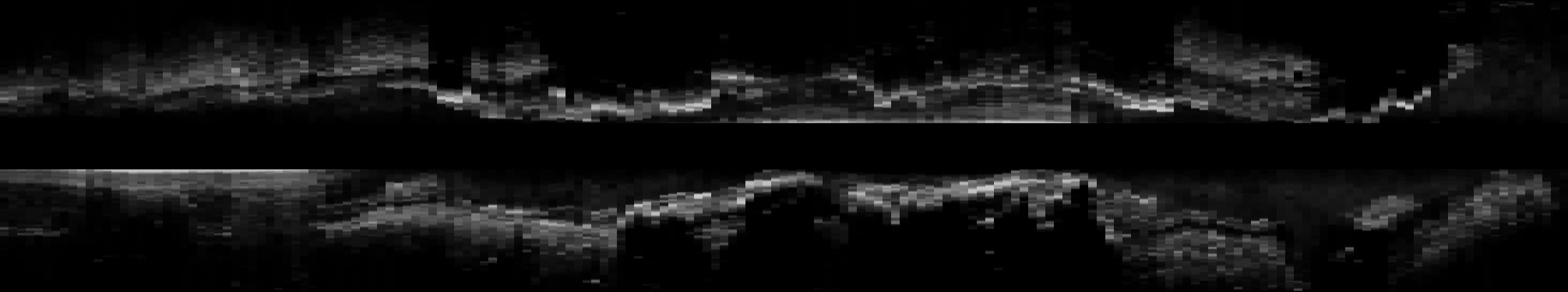}\\
\vspace*{0.05cm}
\includegraphics[width=3.5cm]{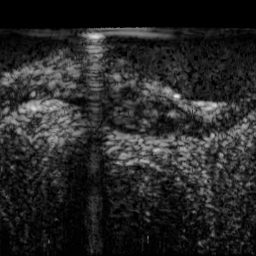}
\includegraphics[width=3.5cm]{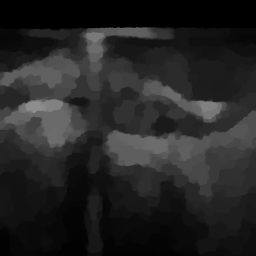}
\includegraphics[width=3.5cm]{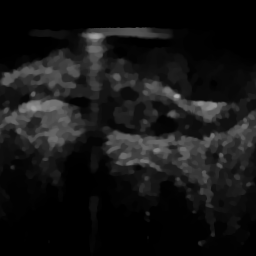}\\
\vspace*{0.05cm}
\includegraphics[width=3.5cm]{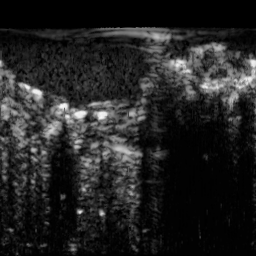}
\includegraphics[width=3.5cm]{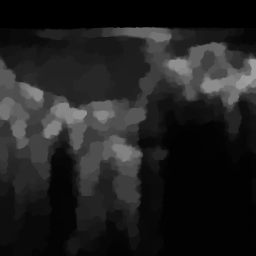}
\includegraphics[width=3.5cm]{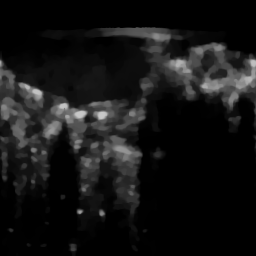}\\
\vspace*{0.05cm}
\includegraphics[width=3.5cm]{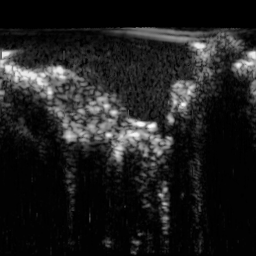}
\includegraphics[width=3.5cm]{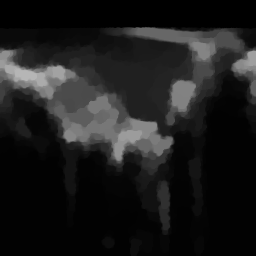}
\includegraphics[width=3.5cm]{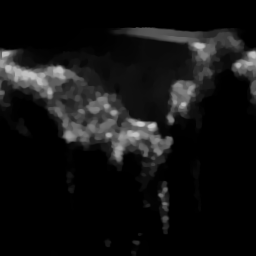}\\
\caption{(Top) Longitudinal views of the original IVUS study, TV-L1 denoised study and MLD denoised study from top to bottom respectively. (Bottom) Cross-sections depicting three anatomical landmarks of interest: (row 1) soft plaque; (row 2) stent site; and (row 3) calcified plaque. Each row presents images of the original IVUS study, TV-L1 denoised study and MLD denoised study from left to right, respectively.}
\label{fig:LViewsInvivo}
\end{figure*}

To evaluate the applicability of the method to medical images, 6 in-vivo IVUS studies were analyzed before and after being denoised by MLD and TV-L1. Each IVUS study underwent gating \cite{MasoTalou2015} to extract the end-diastolic frames, avoiding spatial incoherence in its longitudinal views (see Figure~\ref{fig:LViewsInvivo}). The amount of frames per study after the gating procedure is reported in Table \ref{tab:invivoTVL1vsMLV}. Given the lack of ground truth for denoising on in-vivo dataset, we evaluate the preservation of the most important diagnostic features, i.e., calcifications, stent struts, soft plaque and lumen homogeneization using a qualitative assessment and Pearson correlation against the low-frequencies of the IVUS study. The qualitative analysis is of the utmost importance to ensure that structures of clinical interest are not lost nor degraded during the denoising phase. The correlation analysis renders a quantitative measurement of the structures preserved after denoising.

The sequences were denoised using the parameters described in Table~\ref{tab:invivoTVL1vsMLV}. Note that $\lambda$ was tuned along the studies --by maximizing the Pearson correlation for $\alpha \in \{ 0.1\,i, i=1,2,\ldots,10\}$-- to improve lumen homogeneity and definition of the arterial wall structures for TV-L1 whilst no adjustments were necessary for the MLD across all in-vivo cases. Both methods were iterated until $\lVert J^{k+1} - J^k\rVert_{\infty} < 10^{-5}$ is satisfied. 

The proposed method delivered a higher structural preservations as reported in Table~\ref{tab:invivoTVL1vsMLV} by the low-frequency Pearson correlation. The better correlation is also visualized in the longitudinal views where the intensity bias between the original study and the MLD is lower than with respect to the TV-L1. In general TV-L1 presents intensity saturation of the image, lowering the image contrast and edge sharpness. This saturation and bias in the TV-L1 formulation is expected because its model of the data fidelity term describes a more conservative and biased model (normal distribution with additive noise) with respect to the actual noise distribution (gamma generalized and multiplicative noise).

To outline the improvement of MLD with respect to the classic TV-L1 formulation, we pick three anatomical landmark frames depicting a soft plaque, a stented site and a calcified plaque (respectively first, second and third rows of frames in Figure~\ref{fig:LViewsInvivo}). The soft plaque showed a higher plaque-to-lumen contrast while stair-case artifacts are less frequent --see also the longitudinal view. Hence, plaque details such as the small calcifications in the right part of the image are better described by the MLD image. The stent site is an even better example of the preservation and localization of small structures, as TV-L1 method struggled to preserve them, MLD depicts the struts with the same quality as in the original study. In the bottom row, it is observed that the calcified plaque displays a better calcium-to-plaque contrast and less blurring in the MLD than in the TV-L1. Thus, the use of a better data fidelity term showed to be crucial towards a better restoration of the ultrasonic images. 

\begin{table*}
\centering
\begin{tabular}{l|cc|ccccc}
\toprule
				&	\multicolumn{2}{c|}{TV-L1}						&	\multicolumn{5}{c}{MLD} \\
Study ID (\# of frames)		&	$\alpha$ & $c(I_l,J)$				&	$\alpha$ & $\gamma$ & $\nu$ & $\delta$ & $c(I_l,J)$		\\
\midrule
Case 1 (n=123) 	& $0.3$ & $0.900 \pm 0.026$ & $0.1$ & $1$ & $1$ & $1.1$ & $0.946 \pm 0.008$ \\  
Case 2 (n=111) 	& $0.3$ & $0.890 \pm 0.022$ & $0.1$ & $1$ & $1$ & $1.1$ & $0.951 \pm 0.008$ \\ 
Case 3 (n=171) 	& $0.4$ & $0.932 \pm 0.014$ & $0.1$ & $1$ & $1$ & $1.1$ & $0.951 \pm 0.010$ \\ 
Case 4 (n=68) 	& $0.4$ & $0.948 \pm 0.010$ & $0.1$ & $1$ & $1$ & $1.1$ & $0.959 \pm 0.004$ \\ 
Case 5 (n=183) 	& $0.4$ & $0.929 \pm 0.016$ & $0.1$ & $1$ & $1$ & $1.1$ & $0.954 \pm 0.013$ \\
Case 6 (n=178) 	& $0.4$ & $0.936 \pm 0.014$ & $0.1$ & $1$ & $1$ & $1.1$ & $0.942 \pm 0.018$ \\ 
\bottomrule
\end{tabular}
\vspace*{0.05cm}
\caption{Mean and standard deviation of the Pearson correlation between the low-pass filtered images and the denoised images (with MLD and TV-L1) of 6 in-vivo IVUS studies. The variable $n$ is the amount of frames for each case. The low-pass filtered images, $I_l$, were generated by zeroing the higher half frequencies of its Discrete Fourier Transform.}
\label{tab:invivoTVL1vsMLV}
\end{table*}

\section{Acknowledgements}

This work was partially supported by the Brazilian agencies CNPq, FAPERJ and CAPES. The support of these agencies is gratefully acknowledged.

\section{Conclusions}
		
A novel methodology to specialize the data fidelity term for total variation formulations has been introduced, outlining its advantages for ultrasonic images. The methodology generalizes the construction of the data fidelity term based on the noise and image preprocessing model allowing to easily derive complex case-specific estimators.

The application of this methodology to high-frequency ultrasonic images showed a better contrast, preservation and localization of the structures when compared to using the same formulation with the classic data term. The obtained images depict characteristics better suited for medical image processing such as automatic segmentation, plaque quantification and calcium scoring. 

\bibliographystyle{unsrt}
\bibliography{ms.bib}

\begin{thebibliography}{10}

\bibitem{jain2016survey}
Paras Jain and Vipin Tyagi.
\newblock A survey of edge-preserving image denoising methods.
\newblock {\em Information Systems Frontiers}, 18(1):159--170, 2016.

\bibitem{jain1989fundamentals}
Anil~K Jain.
\newblock {\em Fundamentals of digital image processing}.
\newblock Englewood Cliffs, NJ: Prentice Hall,, 1989.

\bibitem{gonzalez2012digital}
Rafael~C Gonzalez and Richard~E Woods.
\newblock {\em Digital image processing}.
\newblock Upper Saddle River, NJ: Prentice Hall, 2012.

\bibitem{Rudin1992}
Leonid~I. Rudin, Stanley Osher, and Emad Fatemi.
\newblock Nonlinear total variation based noise removal algorithms.
\newblock {\em Physica D: Nonlinear Phenomena}, 60(1-4):259--268, November
  1992.

\bibitem{Chambolle2004}
Antonin Chambolle.
\newblock An algorithm for total variation minimization and applications.
\newblock {\em Journal of Mathematical imaging and vision}, 20(1-2):89--97,
  2004.

\bibitem{gilboa2008nonlocal}
Guy Gilboa and Stanley Osher.
\newblock Nonlocal operators with applications to image processing.
\newblock {\em Multiscale Modeling \& Simulation}, 7(3):1005--1028, 2008.

\bibitem{bredies2010total}
Kristian Bredies, Karl Kunisch, and Thomas Pock.
\newblock Total generalized variation.
\newblock {\em SIAM Journal on Imaging Sciences}, 3(3):492--526, 2010.

\bibitem{tang2016edge}
Liming Tang and Zhuang Fang.
\newblock Edge and contrast preserving in total variation image denoising.
\newblock {\em EURASIP Journal on Advances in Signal Processing}, 2016(1):13,
  2016.

\bibitem{krissian2007oriented}
Karl Krissian, Carl-Fredrik Westin, Ron Kikinis, and Kirby~G Vosburgh.
\newblock Oriented speckle reducing anisotropic diffusion.
\newblock {\em IEEE Transactions on Image Processing}, 16(5):1412--1424, 2007.

\bibitem{perona1990scale}
Pietro Perona and Jitendra Malik.
\newblock Scale-space and edge detection using anisotropic diffusion.
\newblock {\em IEEE Transactions on pattern analysis and machine intelligence},
  12(7):629--639, 1990.

\bibitem{black1998robust}
Michael~J Black, Guillermo Sapiro, David~H Marimont, and David Heeger.
\newblock Robust anisotropic diffusion.
\newblock {\em IEEE Transactions on image processing}, 7(3):421--432, 1998.

\bibitem{weickert1998efficient}
Joachim Weickert, BM~Ter~Haar Romeny, and Max~A Viergever.
\newblock Efficient and reliable schemes for nonlinear diffusion filtering.
\newblock {\em IEEE transactions on image processing}, 7(3):398--410, 1998.

\bibitem{tomasi1998bilateral}
Carlo Tomasi and Roberto Manduchi.
\newblock Bilateral filtering for gray and color images.
\newblock In {\em Computer Vision, 1998. Sixth International Conference on},
  pages 839--846. IEEE, 1998.

\bibitem{durand2002fast}
Fr{\'e}do Durand and Julie Dorsey.
\newblock Fast bilateral filtering for the display of high-dynamic-range
  images.
\newblock In {\em ACM transactions on graphics (TOG)}, volume 21.3, pages
  257--266. ACM, 2002.

\bibitem{gastal2012adaptive}
Eduardo~SL Gastal and Manuel~M Oliveira.
\newblock Adaptive manifolds for real-time high-dimensional filtering.
\newblock {\em ACM Transactions on Graphics (TOG)}, 31(4):33, 2012.

\bibitem{chaudhury2013acceleration}
Kunal~N Chaudhury.
\newblock Acceleration of the shiftable {O}(1) algorithm for bilateral
  filtering and nonlocal means.
\newblock {\em IEEE Transactions on image processing}, 22(4):1291--1300, 2013.

\bibitem{mozerov2015global}
Mikhail~G Mozerov and Joost Van De~Weijer.
\newblock Global color sparseness and a local statistics prior for fast
  bilateral filtering.
\newblock {\em IEEE Transactions on Image Processing}, 24(12):5842--5853, 2015.

\bibitem{buades2005non}
Antoni Buades, Bartomeu Coll, and J-M Morel.
\newblock A non-local algorithm for image denoising.
\newblock In {\em Computer Vision and Pattern Recognition, 2005. CVPR 2005.
  IEEE Computer Society Conference on}, volume~2, pages 60--65. IEEE, 2005.

\bibitem{coupe2008optimized}
Pierrick Coup{\'e}, Pierre Yger, Sylvain Prima, Pierre Hellier, Charles
  Kervrann, and Christian Barillot.
\newblock An optimized blockwise nonlocal means denoising filter for 3-d
  magnetic resonance images.
\newblock {\em IEEE transactions on medical imaging}, 27(4):425--441, 2008.

\bibitem{tasdizen2008principal}
Tolga Tasdizen.
\newblock Principal components for non-local means image denoising.
\newblock In {\em Image Processing, 2008. ICIP 2008. 15th IEEE International
  Conference on}, pages 1728--1731. IEEE, 2008.

\bibitem{maggioni2013nonlocal}
Matteo Maggioni, Vladimir Katkovnik, Karen Egiazarian, and Alessandro Foi.
\newblock Nonlocal transform-domain filter for volumetric data denoising and
  reconstruction.
\newblock {\em IEEE transactions on image processing}, 22(1):119--133, 2013.

\bibitem{donoho1994ideal}
David~L Donoho and Jain~M Johnstone.
\newblock Ideal spatial adaptation by wavelet shrinkage.
\newblock {\em biometrika}, 81(3):425--455, 1994.

\bibitem{donoho1995adapting}
David~L Donoho and Iain~M Johnstone.
\newblock Adapting to unknown smoothness via wavelet shrinkage.
\newblock {\em Journal of the american statistical association},
  90(432):1200--1224, 1995.

\bibitem{chang2000adaptive}
S~Grace Chang, Bin Yu, and Martin Vetterli.
\newblock Adaptive wavelet thresholding for image denoising and compression.
\newblock {\em IEEE transactions on image processing}, 9(9):1532--1546, 2000.

\bibitem{qiu2013llsure}
Tianshuang Qiu, Aiqi Wang, Nannan Yu, and Aimin Song.
\newblock Llsure: local linear sure-based edge-preserving image filtering.
\newblock {\em IEEE Transactions on Image Processing}, 22(1):80--90, 2013.

\bibitem{chambolle1997image}
Antonin Chambolle and Pierre-Louis Lions.
\newblock Image recovery via total variation minimization and related problems.
\newblock {\em Numerische Mathematik}, 76(2):167--188, 1997.

\bibitem{chan2000high}
Tony Chan, Antonio Marquina, and Pep Mulet.
\newblock High-order total variation-based image restoration.
\newblock {\em SIAM Journal on Scientific Computing}, 22(2):503--516, 2000.

\bibitem{kongskov2017directional}
Rasmus~Dalgas Kongskov, Yiqiu Dong, and Kim Knudsen.
\newblock Directional total generalized variation regularization.
\newblock {\em arXiv preprint arXiv:1701.02675}, 2017.

\bibitem{krissian2005speckle}
Karl Krissian, Ron Kikinis, C-F Westin, and Kirby Vosburgh.
\newblock Speckle-constrained filtering of ultrasound images.
\newblock In {\em Computer Vision and Pattern Recognition, 2005. CVPR 2005.
  IEEE Computer Society Conference on}, volume~2, pages 547--552. IEEE, 2005.

\bibitem{le2007variational}
Triet Le, Rick Chartrand, and Thomas~J Asaki.
\newblock A variational approach to reconstructing images corrupted by poisson
  noise.
\newblock {\em Journal of mathematical imaging and vision}, 27(3):257--263,
  2007.

\bibitem{rodriguez2008efficient}
Paul Rodriguez and Brendt Wohlberg.
\newblock An efficient algorithm for sparse representations with lp data
  fidelity term.
\newblock In {\em IEEE Andean Tech. Conf}, 2008.

\bibitem{Michailovich2006}
Oleg~V. Michailovich and Allen Tannenbaum.
\newblock Despeckling of medical ultrasound images.
\newblock {\em Ultrasonics, Ferroelectrics, and Frequency Control, IEEE
  Transactions on}, 53(1):64--78, 2006.

\bibitem{raju2002statistics}
Balasundar~I Raju and Mandayam~A Srinivasan.
\newblock Statistics of envelope of high-frequency ultrasonic backscatter from
  human skin in vivo.
\newblock {\em IEEE transactions on ultrasonics, ferroelectrics, and frequency
  control}, 49(7):871--882, 2002.

\bibitem{rizi2012noise}
F~Yousefi Rizi and SK~Setarehdan.
\newblock Noise reduction in intravascular ultrasound images using curvelet
  transform and adaptive complex diffusion filter: A comparative study.
\newblock In {\em Electrical Engineering (ICEE), 2012 20th Iranian Conference
  on}, pages 1549--1553. IEEE, 2012.

\bibitem{lazrag2012despeckling}
Hassen Lazrag and Med~Saber Naceur.
\newblock Despeckling of intravascular ultrasound images using curvelet
  transform.
\newblock In {\em Sciences of Electronics, Technologies of Information and
  Telecommunications (SETIT), 2012 6th International Conference on}, pages
  365--369. IEEE, 2012.

\bibitem{mitra2015wavelet}
Pabitra Mitra, Chandan Chakraborty, and KM~Mandana.
\newblock Wavelet based non local means filter for despeckling of intravascular
  ultrasound image.
\newblock In {\em Advances in Computing, Communications and Informatics
  (ICACCI), 2015 International Conference on}, pages 1361--1365. IEEE, 2015.

\bibitem{soorajkumar2016fourth}
R~Soorajkumar, P~Krishna Kumar, D~Girish, and Jeny Rajan.
\newblock Fourth order pde based ultrasound despeckling using eni
  classification.
\newblock In {\em Signal Processing and Communications (SPCOM), 2016
  International Conference on}, pages 1--5. IEEE, 2016.

\bibitem{wen2016nonlocal}
Tiexiang Wen, Jia Gu, Ling Li, Wenjian Qin, Lei Wang, and Yaoqin Xie.
\newblock Nonlocal total-variation--based speckle filtering for ultrasound
  images.
\newblock {\em Ultrasonic imaging}, 38(4):254--275, 2016.

\bibitem{Chan1999}
Tony~F. Chan, Gene~H. Golub, and Pep Mulet.
\newblock A {Nonlinear} {Primal}-{Dual} {Method} for {Total}
  {Variation}-{Based} {Image} {Restoration}.
\newblock {\em SIAM Journal on Scientific Computing}, 20(6):1964--1977, January
  1999.

\bibitem{MasoTalou2015}
Gonzalo~Daniel {Maso Talou}, Ignacio Larrabide, Pablo~J Blanco,
  Cristiano~Guedes Bezerra, Pedro~A Lemos, and Ra{\'u}l~A Feij{\'o}o.
\newblock Improving cardiac phase extraction in ivus studies by integration of
  gating methods.
\newblock {\em IEEE Transactions on Biomedical Engineering}, 62(12):2867 --
  2877, 2015.

\end{thebibliography}
	
\appendix	
	\section{Minimum necessary and sufficient conditions} \label{sec:NecSufConditionsAndELEq}
	
	To ease the readability, let us simply denote $I(x,y)$ by $I$. Given the minimization problem presented in Section~\ref{sec:MLD}, stated as
	\begin{equation}
		J = \argmin_{ \tilde{J} } \int_{\Omega} \Big( - c \big( I, \tilde{J} \big)+ \alpha \lvert \nabla \tilde{J} \rvert \Big) \, d\Omega,
		\label{appC:functional}
	\end{equation}
	where 
	\begin{equation}
	 c \big( I, \tilde{J} \big) = \gamma \nu (I- \tilde{J}) - \frac{1}{\delta^{\gamma}} e^{\gamma(I - \tilde{J})},
		\label{appC:MLV}
	\end{equation}
	we have to find the minimizer function ${J} \in \mathcal U$, where
	\begin{equation}
		\mathcal U = \{ \tilde{J} \in \mathbb{R}^{n\times m}; \tilde{J}, \nabla \tilde{J} \mbox{ are square-integrable functions} \}.
	\end{equation} 
	Introducing \eqref{appC:MLV} in \eqref{appC:functional}, we obtain the following functional
	\begin{equation}
	\mathcal F(\tilde J) = \int_{\Omega} \Big( - \gamma \nu (I - \tilde{J}) + \frac{1}{\delta^{\gamma}} e^{\gamma(I - \tilde{J})}+ \alpha \lvert \nabla \tilde{J} \rvert \Big) \, d\Omega.
	\end{equation}
	We use variational calculus to analyze the necessary and sufficient conditions for the function to be a minimizer. Thus, we perturb the function $\tilde{J}$ as $\tilde{J} + \tau \eta$ with $\eta \in \mathcal U$.
	
	The perturbed functional results
	\begin{equation}
	\begin{split}
		\mathcal F(\tilde J + \tau \eta) = \int_{\Omega} &\Big( - \gamma \nu \, (I - \tilde{J}) + \gamma \nu \, \tau \eta \\
			&+ \frac{1}{\delta^{\gamma}} e^{\gamma(I - \tilde{J})} e^{-\gamma \,\tau \eta} + \alpha \lvert \nabla \tilde{J} - \tau \nabla \eta \rvert \Big) \, d\Omega.
	\end{split}
	\end{equation}
	The necessary condition is obtained when the first G\^ateaux derivative of $\mathcal F$ is nullified for any admissible perturbation $\eta$, i.e., $\delta \mathcal F(\tilde J, \eta) = 0, \forall \eta \in \mathcal U$. Then, we calculate such condition as follows
	\begin{multline}
	\delta \mathcal F(\tilde J, \eta) =  \frac{\partial \mathcal F(\tilde J + \tau \eta)}{\partial \tau} \bigg|_{\tau = 0} \\
	= \int_{\Omega} \bigg( \gamma \nu \, \eta - \frac{\gamma}{\delta^{\gamma}} e^{\gamma(I - \tilde{J})} \eta
	+ \alpha \frac{\nabla \tilde{J} \cdot \nabla \eta}{\lvert \nabla \tilde{J} \rvert} \bigg) \, d\Omega = 0\\
    \forall \eta \in \mathcal U.
	\label{appC:devel1stGatDer_1}
	\end{multline}
	
	To determine if the extreme in $\delta \mathcal F(\tilde J, \eta) = 0$ is, in fact, maximum or minimum, we inspect the sign of the second G\^ateaux derivative $\delta^2 \mathcal F(\tilde J, \eta)$ obtained as follows
	\begin{multline}
	\delta^2 \mathcal F(\tilde J, \eta) = \frac{\partial^2 \mathcal F(\tilde J + \tau \eta)}{\partial \tau^2} \bigg|_{\tau = 0} \\
	= \int_{\Omega} \bigg[ \frac{\gamma^2}{\delta^{\gamma}} e^{\gamma(I - \tilde{J})} \eta^2 + \alpha \frac{\nabla \eta \cdot \nabla \eta}{\lvert \nabla \tilde{J} \rvert} \\
 + \alpha \nabla \tilde{J} \cdot \nabla \eta \, \delta \bigg(\frac{1}{\lvert \nabla \tilde{J} + \tau \eta \rvert}\bigg) \bigg] \, d\Omega
	\label{appC:devel2ndGatDer_1}
	\end{multline}
	where 
	\begin{equation}
	\begin{split}
	\delta \bigg(\frac{1}{\lvert \nabla \tilde{J} + \tau \eta \rvert}\bigg) &= - \frac{1}{\lvert \nabla \tilde J \rvert^3} \nabla \tilde J \cdot \nabla \eta.
	\end{split}
	\label{appC:devel2ndGatDer_2}
	\end{equation}
	Replacing \eqref{appC:devel2ndGatDer_2} in \eqref{appC:devel2ndGatDer_1} and rearranging terms
	\begin{multline}
	\delta^2 \mathcal F(\tilde J, \eta) = \int_{\Omega} \bigg[ \frac{\gamma^2}{\delta^{\gamma}} e^{\gamma(I - \tilde{J})} \eta^2 \\
	+ \frac{\alpha}{\lvert \nabla \tilde{J} \rvert} \bigg(\nabla \eta \cdot \nabla \eta - \Big(\frac{\nabla \tilde{J}}{\lvert \nabla \tilde J \rvert} \, \cdot \nabla \eta\Big)^2 \bigg) \bigg] \, d\Omega.
	\label{appC:devel2ndGatDer_3}
	\end{multline}
	Note that the first term of the integral is always positive, and the second term is positive if
	\begin{equation}
	\begin{split}
		\frac{\alpha}{\lvert \nabla \tilde{J} \rvert} \bigg(\nabla \eta \cdot \nabla \eta - \Big(\frac{\nabla \tilde{J}}{\lvert \nabla \tilde J \rvert} \, \cdot \nabla \eta\Big)^2 \bigg) &\ge 0\\
		\nabla \eta \cdot \nabla \eta - \Big(\frac{\nabla \tilde{J}}{\lvert \nabla \tilde J \rvert} \, \cdot \nabla \eta\Big)^2 &\ge 0 \\
		\nabla \eta \cdot \nabla \eta &\ge \Big(\frac{\nabla \tilde{J}}{\lvert \nabla \tilde J \rvert} \, \cdot \nabla \eta\Big)^2  \\
		\lvert \nabla \eta \rvert \, \lvert \nabla \eta \rvert \cos 0 &\ge \bigg( \bigg\lvert \frac{\nabla \tilde{J}}{\lvert \nabla \tilde J \rvert}\bigg\rvert \, \lvert\nabla \eta \rvert \cos \theta\bigg)^2  \\
		\lvert \nabla \eta \rvert^2 &\ge \lvert\nabla \eta \rvert^2 (\cos \theta)^2
	\end{split}
	\end{equation}
	where $\theta$ is the angle between $\frac{\nabla \tilde{J}}{\lvert \nabla \tilde J \rvert}$ and $\nabla \eta$. As $(\cos \theta)^2 \le 1 \Rightarrow \lvert \nabla \eta \rvert^2 \ge \lvert\nabla \eta \rvert^2 (\cos \theta)^2 \Rightarrow$ the second term is always positive. As both terms are always positive, then $\delta^2 \mathcal F(\tilde J, \eta) \ge 0, \forall \eta \in \mathcal U$, implying that $\delta \mathcal F(\tilde J, \eta) = 0$ is a minimum. Note that only $\eta=0$ nullifies the first term since $\gamma \not = 0$ (from the definition of generalized gamma distribution). Also, the second term can be rewritten as
	\begin{equation}
	\begin{split}
		\frac{\alpha}{\lvert \nabla \tilde{J} \rvert} \lvert \nabla \eta \rvert^2 \bigg(1 - (\cos \theta)^2 \bigg)
	\end{split}
	\end{equation}
	where only spatially homogeneous $\eta$ variations or $\nabla \eta \parallel \nabla I$ nullifies such term. Then, both terms are nullified at the same time if and only if $\eta=0$, implying that the minimum is unique.
	
	Finally, we integrate by parts \eqref{appC:devel1stGatDer_1} as follows
	\begin{multline}
	\delta \mathcal F(\tilde J, \eta) = \alpha \frac{\nabla \tilde J}{\lvert \nabla \tilde J \rvert} \cdot \mathbf n \, \eta \bigg|_{\partial \Omega} \\
	+ \int_{\Omega} \bigg[ \gamma \nu - \frac{\gamma}{\delta^{\gamma}} e^{\gamma(I - \tilde{J})} - \alpha \dive\bigg(\frac{\nabla \tilde{J}}{\lvert \nabla \tilde{J} \rvert}\bigg) \bigg] \eta \, d\Omega  = 0\\
    \forall \eta \in \mathcal U,
	\label{appC:devel1stGatDer_2}
	\end{multline}
where $\mathbf n$ is the normal vector to the boundary $\partial \Omega$. By the fundamental lemma of the calculus of variations, we obtain the following Euler-Lagrange equations for \eqref{appC:functional}
	\begin{eqnarray}
		\gamma \nu - \frac{\gamma}{\delta^\gamma} e^{\gamma \, (I - \tilde J)} - \alpha \dive \bigg(\frac{ \nabla \tilde J }{\lvert \nabla \tilde J \rvert}\bigg) = 0& &\mbox{in } \Omega\\ 
		\nabla \tilde J \cdot \mathbf n = 0& &\mbox{on } \partial \Omega.
	\end{eqnarray}
	
	\section{MLD for normally distributed noise} \label{sec:MLDGaussian}
	
	An interesting insight is obtained by deriving the MLD formulation for a normally distributed noise. Let us suppose that $\tilde \varepsilon \sim \mathcal N(\mu,\sigma)$ in \eqref{eq:noiseModelTwoImages}, leading to the probability density function 
	\begin{equation}
	P_{\tilde \varepsilon}(\tilde \varepsilon) = \frac{1}{\sqrt{2 \, \pi \, \sigma^2}} \, e^{-\frac{(\tilde \varepsilon - \mu)^2}{2 \, \sigma^2}}.
	\end{equation}
	Using such distribution, the MLE is derived as follows
	\begin{eqnarray}
		e(I,J) &=& \int_{\Omega} \log \big[ P_{\tilde \varepsilon}(I-J) \big] \nonumber\\
		&=& \int_{\Omega} \log \frac{1}{\sqrt{2 \, \pi \, \sigma^2}} \, e^{-\frac{(I - J - \mu)^2}{2 \, \sigma^2}}\nonumber\\
		&=& C_1 + C_2 \int_{\Omega} -(I - J - \mu)^2 \nonumber\\
		&&\mbox{with } C_1 = - \frac{\lvert \Omega \rvert}{2} \log (2 \, \pi \, \sigma^2), \, C_2 = \frac{1}{2 \, \sigma^2}
		\label{eq:MLEforLogNormal}
	\end{eqnarray}
	As in Section~\ref{sec:MLDGammaNoise}, the constant term and factor are discarded without affecting the MLD outcome, to finally obtain the MLD formulation for normally distributed noise
\begin{equation}
	J = \argmin_{ \tilde{J} } \int_{\Omega} \Big( ( I - \tilde{J} - \mu)^2 + \alpha \lvert \nabla \tilde{J} \rvert \Big) \, d\Omega.
	\label{eq:MLVGaussian}
\end{equation} 
Note that if $\tilde \varepsilon \sim \mathcal N(0,\sigma)$, then the ROF model is recovered \cite{Rudin1992}, which is a specific instantiation of the MLD for zero-mean normally distributed noise.
	
\end{document}